\documentclass[aps,prc,twocolumn,showpacs,preprintnumbers,amsmath,amssymb,superscriptaddress]{revtex4-1}

\usepackage{graphicx}
\DeclareGraphicsExtensions{.pdf,.png,.jpg,.jpeg,.eps}
\usepackage{epsfig}
\usepackage{dcolumn}
\usepackage{bm}
\usepackage{epstopdf}

\begin{document}

\title{$^{137,138,139}$La($n$, $\gamma$) cross sections constrained with statistical decay properties of $^{138,139,140}$La nuclei}

\author{B.~V.~Kheswa}
\affiliation{Department of Physics, University of Oslo, N-0316 Oslo, Norway}
\affiliation{Physics Department, University of Stellenbosch, Private Bag X1, Matieland 7602, Stellenbosch, South Africa}

\author{M.~Wiedeking}
\affiliation{iThemba LABS, P.O. Box 722, 7129 Somerset West, South Africa}
\author{J.~A.~Brown}
\affiliation{Department of Nuclear Engineering, University of California, Berkeley, California 94720, USA}
\author{A.~C.~Larsen}
\affiliation{Department of Physics, University of Oslo, N-0316 Oslo, Norway}
\author{S.~Goriely}
\affiliation{Institut d'Astronomie et d'Astrophysique, Universit\'e Libre de Bruxelles, CP 226, B-1050 Brussels, Belgium}
\author{M.~Guttormsen}
\affiliation{Department of Physics, University of Oslo, N-0316 Oslo, Norway}
\author{F.~L.~Bello Garrote}
\affiliation{Department of Physics, University of Oslo, N-0316 Oslo, Norway}
\author{L.~A.~Bernstein}
\affiliation{Department of Nuclear Engineering, University of California, Berkeley, California 94720, USA}
\affiliation{Nuclear Science Division, Lawrence Berkeley National Laboratory, Berkeley, California 94720, USA}
\affiliation{Physical and Life Sciences Directorate, Lawrence Livermore National Laboratory, Livermore, California 94551, USA}
\author{D.~L.~Bleuel}
\affiliation{Physical and Life Sciences Directorate, Lawrence Livermore National Laboratory, Livermore, California 94551, USA}
\author{T.~K.~Eriksen}
\affiliation{Department of Physics, University of Oslo, N-0316 Oslo, Norway}
\author{F.~Giacoppo}
\affiliation{Helmholtz Institute Mainz, 55099 Mainz, Germany}
\affiliation{GSI Helmholtzzentrum für Schwerionenforschung, 64291 Darmstadt, Germany}
\author{A.~G\"orgen}
\affiliation{Department of Physics, University of Oslo, N-0316 Oslo, Norway}
\author{B.~L.~Goldblum}
\affiliation{Department of Nuclear Engineering, University of California, Berkeley, California 94720, USA}
\author{T.~W.~Hagen}
\affiliation{Department of Physics, University of Oslo, N-0316 Oslo, Norway}
\author{P.~E.~Koehler}
\affiliation{Department of Physics, University of Oslo, N-0316 Oslo, Norway}
\author{M.~Klintefjord}
\affiliation{Department of Physics, University of Oslo, N-0316 Oslo, Norway}

\author{K. L.~Malatji}
\affiliation{iThemba LABS, P.O. Box 722, 7129 Somerset West, South Africa}
\affiliation{Physics Department, University of Stellenbosch, Private Bag X1, Matieland 7602, Stellenbosch, South Africa}

\author{J. E.~Midtb{\o}}
\affiliation{Department of Physics, University of Oslo, N-0316 Oslo, Norway}
\author{H.~T.~Nyhus}
\affiliation{Department of Physics, University of Oslo, N-0316 Oslo, Norway}
\author{P.~Papka}
\affiliation{Physics Department, University of Stellenbosch, Private Bag X1, Matieland 7602, Stellenbosch, South Africa}
\author{T.~Renstr{\o}m}
\affiliation{Department of Physics, University of Oslo, N-0316 Oslo, Norway}
\author{S.~J.~Rose}
\affiliation{Department of Physics, University of Oslo, N-0316 Oslo, Norway}
\author{E.~Sahin}
\affiliation{Department of Physics, University of Oslo, N-0316 Oslo, Norway}
\author{S.~Siem}
\affiliation{Department of Physics, University of Oslo, N-0316 Oslo, Norway}
\author{T.~G.~Tornyi}
\affiliation{Department of Physics, University of Oslo, N-0316 Oslo, Norway}

\date{\today}

\begin{abstract}

The nuclear level densities and $\gamma$-ray strength functions of $^{138,139,140}$La were measured using the $^{139}$La($^{3}$He, $\alpha$), $^{139}$La($^{3}$He, $^{3}$He$^\prime$)  and  $^{139}$La(d, p) reactions. The particle-$\gamma$ coincidences were recorded with the silicon particle telescope (SiRi) and NaI(Tl) (CACTUS) arrays. In the context of these experimental results, the low-energy enhancement in the A$\sim$140 region is discussed. The $^{137,138,139}$La($n, \gamma)$ cross sections were calculated at $s$- and $p$-process temperatures using the experimentally measured nuclear level densities and $\gamma$-ray strength functions. Good agreement is found between $^{139}$La($n, \gamma)$ calculated cross sections and previous measurements. 

\end{abstract}
\vspace{0.5cm}


\maketitle

\section{Introduction}

At relatively low excitation energies, $E_{x}$, well resolved quantum states are available to which a nucleus can be excited. The $E_{x}$, spins and parities ($J^{\pi}$) of these states, as well as the electromagnetic properties of $\gamma$-ray transitions can be measured using standard particle and $\gamma$-ray spectroscopic techniques. In contrast, as $E_{x}$ approaches the neutron separation energy ($S_{n}$) the number and widths of levels increases dramatically and create a quasi-continuum. In this region states cannot be resolved individually to measure their decay properties. Instead of using discrete spectroscopic tools, a broad range of techniques has been developed to extract statistical properties, below or in the vicinity of $S_{n}$, such as the nuclear level density (NLD) and $\gamma$-ray strength function ($\gamma$SF) which are  measures of the average nuclear response. Some of the commonly used experimental methods include (i) ($\gamma$,$\gamma$') scattering  using mono-energetic beams \cite{Tonchev2010, Angell2012} or Bremsstrahlung photon sources \protect\cite{Schwengner09, Ozel2014}, (ii) ($n, \gamma$) measurements with thermal/cold neutron beams \cite{Firestone2007, Kudejova2000}, average resonance capture \cite{arc1}, (iii) two-step cascade methods using thermal neutrons \cite{Becvar92} or charged particle reactions \cite{Wiedeking12}, and (iv) isoscalar sensitive techniques \cite{savran2013, krumbholz2015, pellegri2014, negi2016}. 

At the University of Oslo a powerful experimental method, known as the Oslo Method \cite{AS00}, was developed. It is based on charged particle-$\gamma$ coincidence data from scattering or transfer reactions and allows for the simultaneous extraction of the NLD and $\gamma$SF up to $S_{n}$.  The $\gamma$SF extracted with the Oslo Method can not only be used to identify and enhance our understanding of resonance structures on the low-energy tail of the giant electric dipole resonance, but also to obtain sensitive nuclear structure information such as the $\gamma$ deformation from scissors resonances \cite{Guttormsen2014,Laplace2016}. The $\gamma$SF has the potential to significantly impact reaction cross sections and therefore astrophysical element formation \protect\cite{Arn2003,Arn2007} and advanced nuclear fuel cycles \protect\cite{AFC}. Measurements of the NLD provides insight into the evolution of the density of states for different nuclei \cite{GuttormsenNLD} and can be used to determine nuclear thermodynamic properties such as entropy, nuclear temperature, and heat capacity as a function of $E_{x}$ \cite{Gia2014, Moretto2015}. 

In the present paper we report on the details of the NLDs and $\gamma$SFs, extracted using the Oslo Method, of $^{138,139,140}$La and the corresponding ($n$, $\gamma$) cross sections. The $^{138,139}$La experimental results have already been used to investigate the synthesis of $^{138}$La in $p$-process environments \cite{BVK2015} and were able to reduce the uncertainties of its production significantly. The findings do not favour the $^{138}$La production by photodisintegration processes, but rather the theory that $^{138}$La is produced through neutrino-induced reactions \cite{Woosley1990, Kajino2014}, with the $\nu_e$-capture on $^{138}$Ba as the largest contributor \cite{Goriely2001, Byelikov2007}.

\section{Experimental Details}

Two experiments were performed at the cyclotron laboratory of the University of Oslo, over two consecutive weeks, with a 2.5 mg/cm$^{2}$ thick natural $^{139}$La target and $^{3}$He and deuterium beams. The excited $^{138,139}$La nuclei were produced through the $^{139}$La($^{3}$He, $\alpha$) and $^{139}$La($^{3}$He, $^{3}$He$^\prime$) reaction channels at a beam energy of 38 MeV, while $^{140}$La was obtained from $^{139}$La(d, p) reactions at 13.5 MeV beam energy. The $\alpha$-$\gamma$, $^{3}$He-$\gamma$ and $p$-$\gamma$ coincident events were detected with the SiRi \cite{a} and CACTUS \cite{b} arrays within a 3 $\mu$s time window and recorded. During the offline analysis the time gate was decreased to 50 ns for $^{138,139}$La and 40 ns for $^{140}$La. The SiRi array consists of 64 $\Delta$E-E Si detector telescopes (130 and 1550 $\mu$m thick $\Delta$E and E, respectively) and was positioned 50 mm from the target at $\theta_{Lab}$ = 47$^{\circ}$ with respect to the beam axis, covering a total solid angle of $\approx$ 6$\%$. CACTUS comprised 26 collimated 5$^{\prime\prime}$x5$^{\prime\prime}$ NaI(Tl) detectors mounted on a spherical frame, enclosing the target located at the center, with a total efficiency of 14.1$\%$ for 1.3 MeV $\gamma$-ray transitions. 

\begin{figure}[h!]
\centering
\includegraphics[scale=0.44]{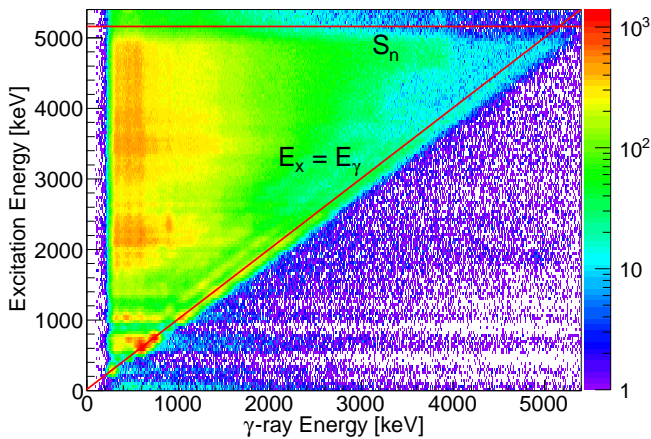}
\caption{(color online) The $E_{x}$ vs $E_{\gamma}$ matrix for $^{140}$La. The 45$^{\circ}$ diagonal line is intended to guide the eye and shows the location of one-step decays to the 3$^{-}$ ground state of $^{140}$La. The neutron separation energy, $S_{n}$, is indicated by the horizontal red line. This comprises the raw $\gamma$ spectra before unfolding.}
\label{matrixa}
\end{figure}

The measured $\alpha$, $^{3}$He and $p$ energies were converted to $E_{x}$ for each of the compound nuclei $^{138,139,140}$La. Kinematic corrections due to the geometry of the setup and the Q-values of 11800 and 2936 keV \cite{NNDC} of the respective reactions ($^{3}$He, $\alpha$) and ($d$, $p$) were taken into account. A typical $E_{x}$ vs $E_{\gamma}$ matrix is shown in Fig. \ref{matrixa} for $^{140}$La, and similar matrices were extracted for $^{138,139}$La. Above $S_{n}$ there is a significant decrease in the number of events due to the dominating neutron emission probability.     

\section{Oslo Method}

A brief outline of the analytical methodology is given here, but a more detailed description of the Oslo Method can be found in Ref. \cite{AS00}. The $\gamma$-ray spectra of $^{138,139,140}$La nuclei were unfolded using the detector response functions and iterative unfolding method \cite{e}. Thus the contributions from pair production and Compton scattering were eliminated and only the true full-energy spectra were obtained. From these, the primary $\gamma$-ray spectra were extracted according to the first generation method \cite{f}.

The $\gamma$SF and NLD of all three La isotopes were extracted from the corresponding primary $\gamma$-ray matrices, $P(E_{\gamma}, E_{x})$, referred to as the first-generation matrices \cite{AS00}. According to Fermi$^\prime$s golden rule \cite{Dirac1927,Fermi1950}, the probability of decay from an initial state $i$ to a set of final states $j$ is proportional to the level density at the final state, $\rho(E_{f})$ where $E_{f} = E_{i} - E_{\gamma}$, and the transition matrix element, $|\langle f|H'|i \rangle|^{2}$. The first-generation matrix is proportional to the $\gamma$-ray decay probability and can be factorized according to Fermi$'$s golden rule equivalent expression

\begin{equation}
\centering
P(E_{\gamma}, E_{x}) \propto \rho(E_{f}) \mathcal{T}_{if},
\label{fact_equation}
\end{equation}

\noindent where $\mathcal{T}_{if}$ is a $\gamma$-ray transmission coefficient for the decay from state $i$ to state $f$. Assuming the validity of the Brink hypothesis \cite{w} and generalizing it to any collective excitation implies that $\mathcal{T}_{if}$ is only dependent on the $\gamma$-ray energy ($E_{\gamma}$) and not on the properties of the states $i$ and $f$ and equations \eqref{fact_equation} becomes 

\begin{equation}
\centering
P(E_{\gamma}, E_{x}) \propto \rho(E_{f}) \mathcal{T}(E_\gamma). 
\label{fact_equationa}
\end{equation}

The $\mathcal{T}(E_{\gamma})$  and $\rho(E_{f})$ are simultaneously extracted by fitting the theoretical first generation matrix $P_{th}(E_{\gamma}, E_{x})$ to the experimental $P(E_{\gamma}, E_{x})$ according to \cite{AS00}


\begin{equation}
\label{eq:eq1a}
\chi^2=\frac{1}{N}\sum_{E_{x}}\sum_{E_{\gamma}} \left( \frac{P_{th}(E_{\gamma}, E_{x})-P(E_{\gamma}, E_{x})}{\varDelta P(E_{\gamma}, E_{x})}  \right)^{2},
\end{equation}

\noindent where $N$ and $\varDelta P(E_{\gamma}, E_{x})$ are the degrees of freedom and the uncertainty in the primary matrix, respectively. The theoretical first-generation matrix can be estimated from

\begin{equation}
P_{th} (E_{\gamma}, E_{x})=\frac{\rho(E_{f}) \mathcal{T}(E_{\gamma})} { {\sum\limits_{E_{\gamma}}} \rho({E_{f}}) \mathcal{T}({E_{\gamma}})}.
\end{equation}

\begin{figure*}
\centering
\includegraphics[scale=0.94]{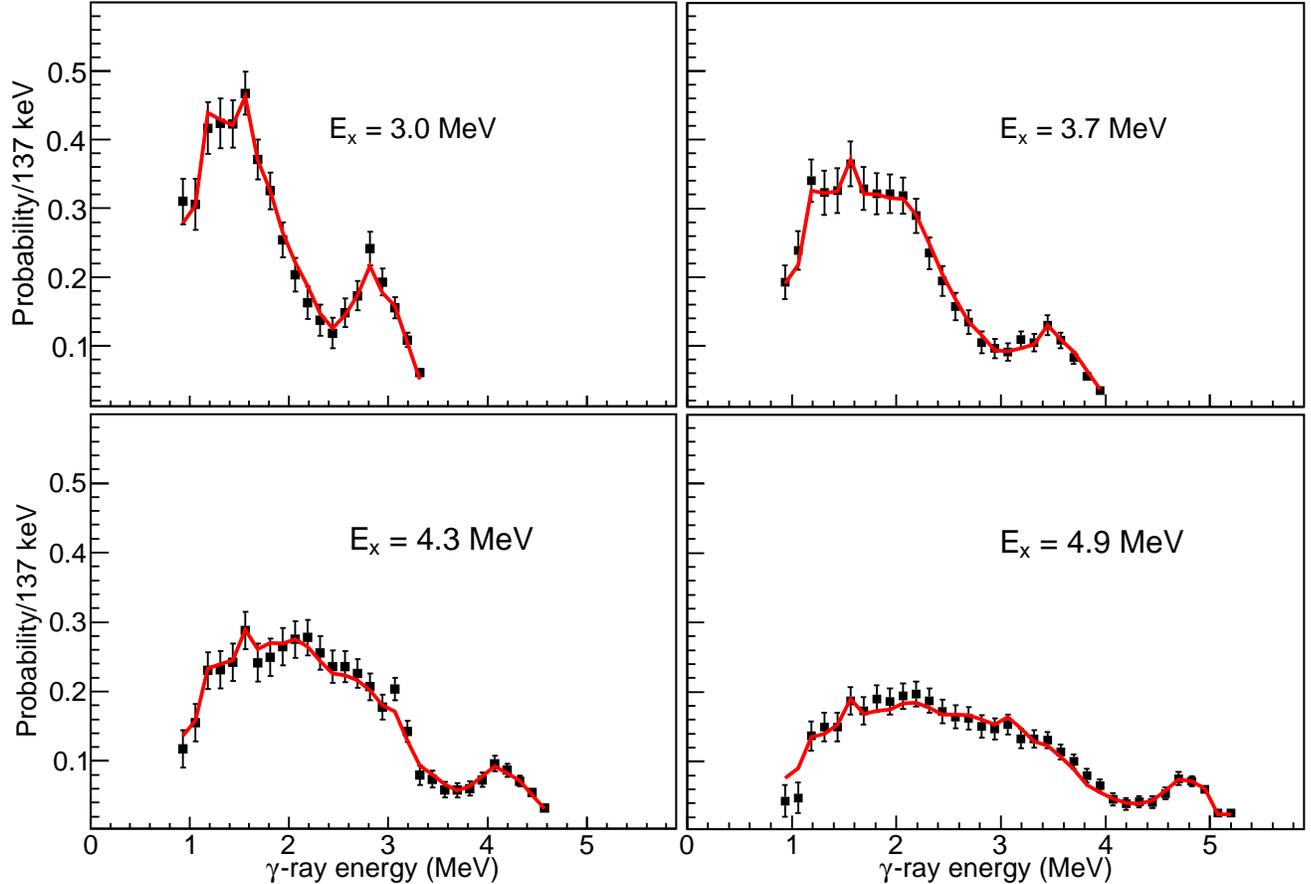}
\caption{(color online) The goodness-of-fit between first-generation matrices for $^{140}$La. The calculated $P_{th} (E_{x}, E_{\gamma})$ (red curve) and experimental $P(E_{x}, E_{\gamma})$ (black data points) at different excitation energies, $E_x$.}
\label{goodness_of_fitc}
\end{figure*}

The $\chi^2$ minimization was performed in the energy regions of 1 MeV $\leq$ $E_{\gamma}$ $\leq$ 7.1 MeV and 3.5 MeV $\leq$ $E_{x}$ $\leq$ 7.1 MeV for $^{138}$La, 1.7 MeV $\leq$ $E_{\gamma}$ $\leq$ 8.5 MeV and 3.5 MeV $\leq$ $E_{x}$ $\leq$ 8.5 MeV for $^{139}$La, and 1 MeV $\leq$ $E_{\gamma}$ $\leq$ 5 MeV and 2.8 MeV $\leq$ $E_{x}$ $\leq$ 5 MeV for $^{140}$La. The ranges were determined by inspection of the matrices and exclude non-statistical structures. The goodness of fit between $P(E_{\gamma}, E_{x})$ and $P_{th}(E_{\gamma}, E_{x})$ is illustrated for $^{140}$La, at various bins of $E_{x}$, in Fig. \ref{goodness_of_fitc}. This comparison is equally good for all spectra and demonstrates the excellent agreement between the theoretical and experimental first-generation matrices. Hence it allows for the extraction of the correct $\rho(E_{f})$ and $\mathcal{T}(E_\gamma)$. Similar fits are also obtained for $^{138,139}$La.

\section{Results}

The procedure outlined in Sec. III yields a functional form for $\rho(E_{f})$ and $\mathcal{T}(E_{\gamma})$ which must be normalized to known experimental data to obtain  physical solutions. It can be shown that infinitely many solutions of Eq. \eqref{eq:eq1a} can be obtained and expressed in the form \cite{AS00}:

\begin{equation}
\tilde{\rho}(E_{f}) = A\rho(E_{f})e^{\alpha E_{f}}
\label{q1}
\end{equation}
\begin{equation}
\tilde{\mathcal{T}}(E_{\gamma})= B\mathcal{T}(E_\gamma)e^{\alpha E_{\gamma}},
\label{q2}
\end{equation}

\noindent where the $\alpha$ parameter is the common slope between $\tilde{\rho}(E_{f})$  and $\tilde{\mathcal{T}}(E_{\gamma})$ and $A$, $B$ are normalization parameters. The values of $\alpha$ and $A$ are obtained by normalizing $\tilde{\rho}(E_{f})$ to $\rho(S_{n})$ and to the level density of known discrete states. 

\subsection{Nuclear Level Densities}

Two theoretical models were used to obtain different values of $\rho(S_{n})$  for each isotope. These are the i) Hartree-Fock-Bogoliubov + Combinatorial (HFB + Comb.) \cite{h} and ii) Constant Temperature + Fermi Gas (CT + FG) model with both parities assumed to have equal contributions. In the latter case, two spin cut-off parameter prescriptions were considered. Thus we explored three different normalizations for each La isotope.  

The HFB + Comb. model is a microscopic combinatorial approach that is used to calculate an energy-, spin-, and parity-dependent NLD. It uses the HFB single-particle level scheme to compute incoherent particle-hole state densities as a function of $E_{x}$, spin projection on the intrinsic symmetry axis of the nucleus, and parity. Once the incoherent state densities have been determined, the collective effects such as rotational and vibrational enhancement are accounted for. As shown in Ref. \cite{h}, these microscopic NLDs can be further normalized to reproduce the experimental neutron resonance spacing at $S_{n}$, hence determining $\rho(S_{n})$, and to the level density of known discrete states. 

The first normalization with the CT + FG model is based on the spin cut-off parameter of Ref. \cite{Edigy2005} and we calculate $\rho(S_{n})$ according to \cite{AS00}:

\begin{equation}
\rho({S_{n}})=\frac{2\sigma^{2}}{D_{0}(J_{T}+1)e^{[{-(J_{T}+1)^{2}}/{2\sigma^{2}}]} + e^{\left({-J^{2}_{T}}/{2\sigma^{2}}\right)}J_{T}},
\label{BSFGa}
\end{equation} 	

\noindent where $D_{0}$, $\sigma$, and $J_{T}$ are the $s$-wave resonance spacing, spin cut-off parameter, and spin of a target nucleus in $(n, \gamma)$ reactions. The spin cut-off parameter is given by \cite{Edigy2005}:

\begin{equation}
\sigma^2 = 0.0146A^{\frac{5}{3}}\frac{  \sqrt{1 + 4a(E_{x}- E_{1})}  }{2a},
\label{cutoff}
\end{equation} 

\noindent where $a$, $E_{1}$ and $A$ are level density parameter, excitation energy shift and nuclear mass. In addition to $\rho(S_n)$, the NLD for other $E_{x}$ regions was computed with the constant temperature law \cite{TEricson1959}: 

\begin{equation}
\rho(E_{x}) = \frac{1}{T}e^{\frac{E_{x}-E_{0}}{T}},
\label{eqq4}
\end{equation}

\noindent where $T$ and $E_{0}$ are the nuclear temperature and energy-shift parameter, respectively. The FG spin distribution was assumed for all $E_{x}$. 

In the second approach, $\rho(E_{x}, J)$ was calculated with the spin cut-off parameter equation as implemented in the TALYS code \cite{Kon2008}. Here the excitation energy is divided into two regions separated by the matching energy $E_{M}$, the point where values from different models and their derivatives are equal. For 0 $<$ $E_{x}$ $<$ $E_{M}$ the Constant Temperature (CT) model is used, while for $E_{x}$ $>$ $E_{M}$, including $S_{n}$, the FG model is used: 

\begin{equation}
\rho(E_{x}) = \frac{1}{12 \sigma \sqrt{2}} \frac{e^{2 \sqrt{a(E_{x}-\delta)}}}{a^{\frac{1}{4}}(E_{x}-\delta)^{\frac{5}{4}}}, 
\label{ldm1}
\end{equation} 

\noindent where $a$ and $\sigma$ are the level density parameter and width of the spin distribution, respectively. The energy $\delta$ accounts for breaking of nucleon pairs that is required before the excitation of individual components. The spin cut-off parameter at $S_{n}$ was calculated from TALYS with \cite{Kon2008}:

\begin{equation}
\sigma^2 = 0.01389A^{\frac{5}{3}} \frac{\sqrt{a(E_{x}-\delta)}}{\bar{a}}, 
\label{fgspina}
\end{equation} 


\noindent where $\bar{a}$ is the asymptotic level density parameter that would be obtained in the absence of any shell effect. For the remainder of this contribution we refer to the CT + FG model that is based on Eq. (8) as the BSFG1 + CT, and that from Eq. (11) as the BSFG2 + CT model. 

\begin{figure}[h!]
\centering
\includegraphics[scale=0.510]{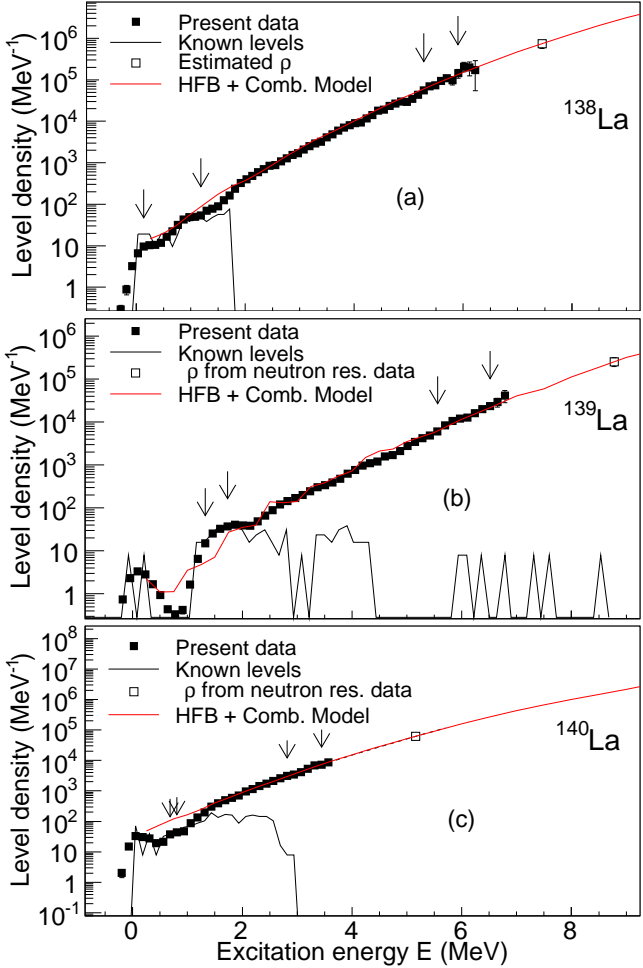}
\caption{(color online) The experimental NLD (black data) of $^{138}$La (a), $^{139}$La (b), and $^{140}$La (c), and the microscopic calculated (red line) $\rho(E_{x})$. The solid black lines are the level densities of know discrete states, while the sets of vertical arrows at low and high energies show regions where the experimental $\rho(E_{x})$ was normalized to the level density of known discrete states and $\rho(S_{n})$.}
\label{nld_La}
\end{figure}

\begin{figure}[h!]
	\centering
	\includegraphics[scale=0.510]{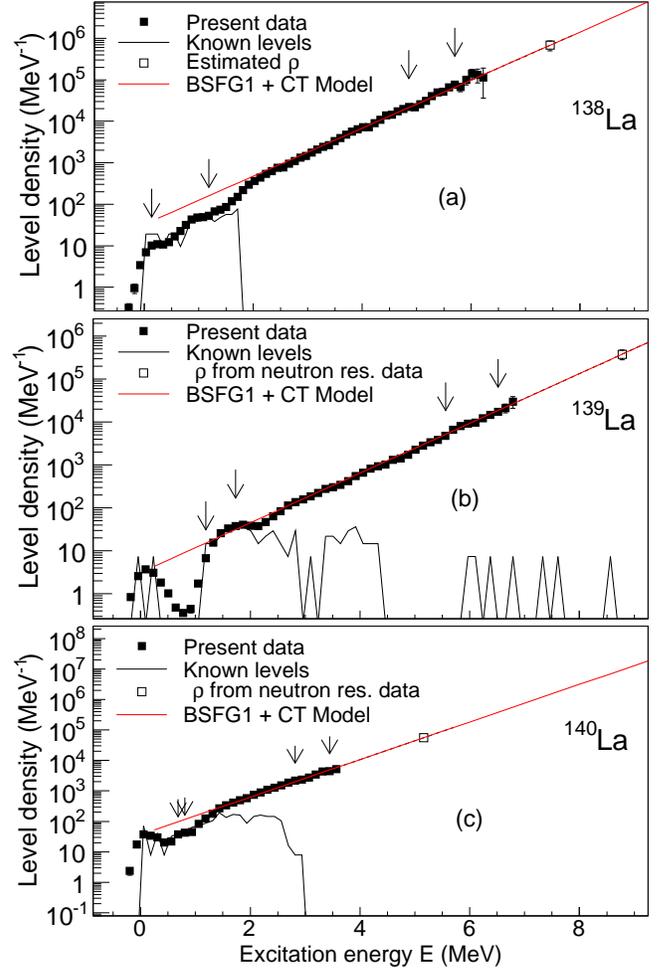}
	\caption{(color online) The NLD (black data) normalized using the Fermi gas model based on Eq. \eqref{cutoff}, for the three La nuclei. The red line shows the CT model used for extrapolation of level density. The solid black lines are the level densities of know discrete states, while the sets of vertical arrows at low and high energies show regions where the experimental $\rho(E_{x})$ were normalized to the level density of known discrete states and $\rho(S_{n})$.}
	\label{nld_LaC}
\end{figure}

\begin{figure}[h!]
	\centering
	\includegraphics[scale=0.48]{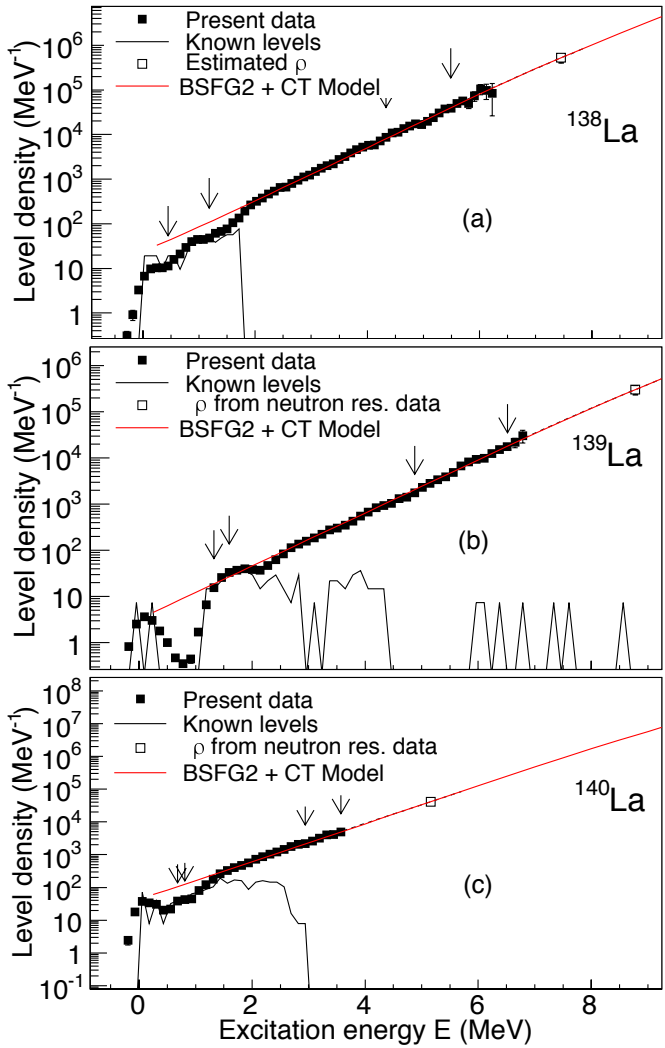}
	\caption{(color online) NLD(black data) normalized to $\rho(S_{n})$ obtained with the Fermi gas model as implemented in TALYS \cite{Kon2008}. The solid black lines are the level densities of know discrete states, while the sets of vertical arrows at low and high energies show regions where the experimental $\rho(E_{x})$ was normalized to the level density of known discrete states and $\rho(S_{n})$.}
	\label{nld_LaB}
\end{figure}

The normalized $\rho(E_{x})$ from models HFB+Comb, BSFG1 + CT, and  BSFG2 + CT are shown in Figs. \ref{nld_La}, \ref{nld_LaC}, and \ref{nld_LaB}, respectively. In each figure these $\rho(E_{x})$ are superimposed with their corresponding theoretical NLDs for comparison. In the case of $^{138}$La there is no $D_{0}$ measurements from ($n$, $\gamma$) resonance experiments due to the unavailability of $^{137}$La target material. Hence, we used the estimated value which was taken from our previous work \cite{BVK2015}. Similarly the experimental average radiative width $\langle   \varGamma_{\gamma}(S_{n},  J_{T}, \pi_{T})  \rangle$, used for the normalization,  was estimated with a spline fit as implemented in the TALYS reaction code. For $^{139}$La, $D_{0}$ and $\langle   \varGamma_{\gamma}(S_{n},  J_{T}, \pi_{T})  \rangle$ are averages of experimental values taken from \cite{p, q}, while for $^{140}$La they were obtained from Ref. \cite{p} only. The experimental NLD does not reach energies above $S_{n}$ - $E_{\gamma}^{min}$, where $E_{\gamma}^{min}$ is the minimum $\gamma$-ray energy considered in the extraction of the $\gamma$SF and $\rho(E_{x})$, as discussed in Sec. III. As a result, the interpolation between experimental data to $\rho(S_{n})$ is accomplished using the models discussed (see Figs. \ref{nld_La}, \ref{nld_LaC} and \ref{nld_LaB}). The normalization parameters for the three La isotopes are provided in Tab. \ref{table1}.

\begin{table*}[htb]
	
	\caption{Structure data and normalization parameters for $^{138,139,140}$La.} 
	\begin{tabular}{lccccccccc}
		\hline
		\hline
		Nucleus    & $I_t^\pi$ &       $D_0$   &       $S_n$  & $\sigma_{BSFG2}(S_n)$    & $\sigma_{BSFG1}(S_n)$  & $\rho_{BSFG2}(S_n)$  &   $\rho_{BSFG1}(S_n)$   &     $\rho_{HFB}(S_n)$    & $\langle   \varGamma_{\gamma}(S_{n},  J_{T}, \pi_{T})  \rangle$    \\
		&            &                 [eV]        &     [MeV]        &                 &         &         [$10^4$ MeV$^{-1}$]&    [$10^4$ MeV$^{-1}$]&    [$10^4$ MeV$^{-1}$]   &  [meV]                  \\
		\hline
		$^{138}$La &  $7/2^+$     & 20.0 $\pm$4.4$^{a}$      &  7.452  &    5.7$\pm$0.6 &  6.7$\pm$0.7  & 52.3$\pm$12  & 68.1$\pm$18.6&    74.2$\pm$ 17.0    &71.0$\pm$13.6$^{b}$  \\
		
		$^{139}$La &  $5^+$   &  31.8$\pm$7.0    & 8.778      &   5.8$\pm$0.6   &  6.9$\pm$0.7     & 30.1$\pm$7.0 &    37.8$\pm$9.7     &    25.5 $\pm$ 7.0      &95.0$\pm$18.2     \\
		$^{140}$La &  $7/2^+$     & 220$\pm$20         & 5.161  &   5.0$\pm$0.5  &  6.2$\pm$0.6     & 4.1$\pm$0.4 & 5.5$\pm$1.0 &     6.2 $\pm$ 0.7                   & 55.0$\pm$2.0  \\
		\hline 
		\hline
	\end{tabular}
	\\
	\label{tab:nldpar}
	$^a$ Estimated (see Ref. \cite{BVK2015} for details).\\
	$^{b}$ Estimated with the spline fit that is implemented in the TALYS reaction code.
	\label{table1}
\end{table*}

\subsection{$\gamma$-ray Strength Function}

With the assumption that statistical decays of the residual nuclei are dominated by dipole transitions \cite{LarsenPRL2013}, the $\gamma$SF can be calculated from the $\gamma$-ray transmission coefficient according to:

\begin{equation}
f(E_{\gamma}) = \frac{B\mathcal{T} (E_{\gamma}) e^{\alpha E_{\gamma}}}{2 \pi E_{\gamma}^{3}}.
\label{eqq5}
\end{equation}

\begin{figure}[h!]
	\centering
	\includegraphics[scale=0.48]{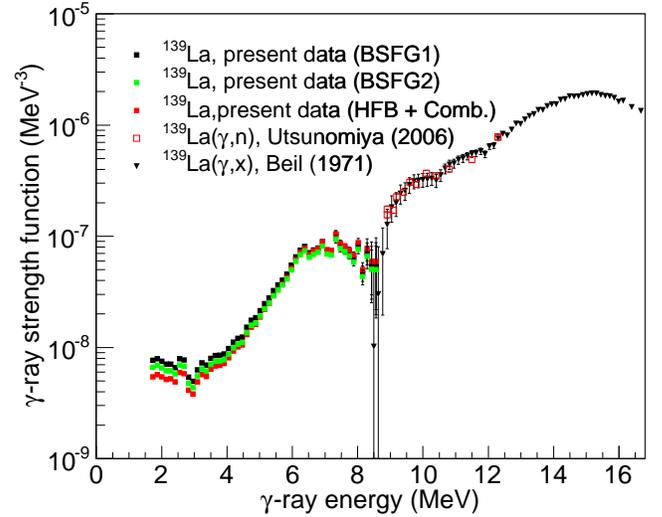}
	\caption{(color online) The $\gamma$SF of $^{139}$La, normalized using spin distributions obtained within HFB + Comb.(red data) and Fermi gas (BSFG1 + CT (black data) and BSFG2 + CT (green data)) models, and compared with photo-neutron data \cite{Utsunomiya2006, Beil1971}.}
	\label{gsf139}
\end{figure}

\begin{figure}[h!]
	\centering
	\includegraphics[scale=0.48]{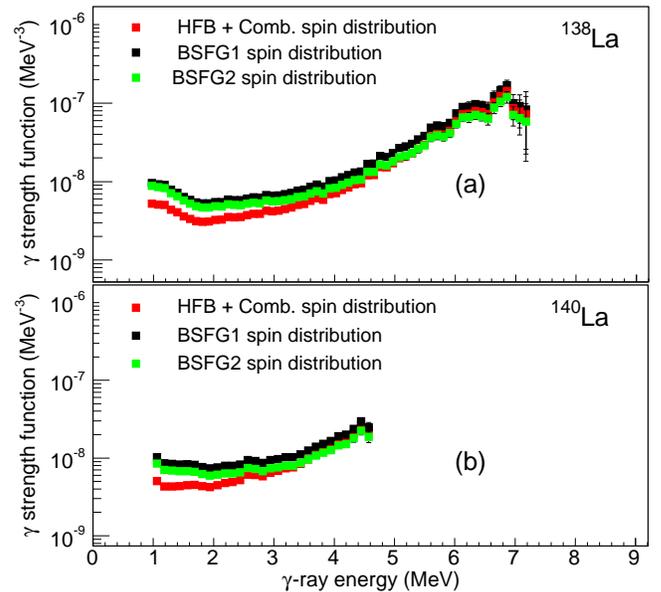}
	\caption{(color online) The $\gamma$SF of $^{138,140}$La (panels (a) and (b)), normalized using spin distributions from Fermi gas (Eq. \eqref{cutoff}  (black data) and Eq. \eqref{fgspina} (green data)) and HFB + Comb. (red data) models.}
	\label{gsf}
\end{figure}

\noindent The absolute normalization parameter $B$ is calculated from  $\langle   \varGamma_{\gamma}(S_{n},  J_{T}, \pi_{T})  \rangle$ according to \cite{JKopecky1941}:

\begin{equation}
\begin{split}
\langle   \varGamma_{\gamma}(S_{n},  J_{T}, \pi_{T})  \rangle =\frac{B}{4\pi D_{0}} \int_{0}^{S_n} \mathcal{T}(E_\gamma)  \rho(S_{n}-E_{\gamma}) dE_{\gamma}\\
\times\sum_{J=-1}^{1} g(S_{n}-E_{\gamma}, J_{T}\pm\frac{1}{2}+J),
\end{split}
\label{q9}
\end{equation}  

\noindent where $J_{T}$ and $\pi_{T}$ are the spin and parity of the target nucleus in the ($n$, $\gamma$) reaction, and $\rho(S_{n} - E_{\gamma} )$ is the experimental level density. The spin distributions $g(E_{x}$,$J)$ were assumed to follow Gaussian distributions with energy-dependent $\sigma$ which were obtained separately from the HFB + Comb., BSFG1 + CT, and BSFG2 + CT models. These were normalized such that $\sum_{J}^{}g(E_{x}, J) \approx 1$. The $\gamma$SF normalized with all three spin distributions are individually compared for each La isotope in Figs. \ref{gsf139} and \ref{gsf}. For  $^{139}$La these are further compared to the giant electric dipole resonance data taken from  \cite{Utsunomiya2006, Beil1971}. The normalization parameters for the three La isotopes are provided in Table \ref{table1}.

\section{Discussion}

For $^{138,140}$La our measurements provide the first data of the $\gamma$SF and NLD below $S_n$. For $^{139}$La data are available from $(\gamma,\gamma^\prime)$ measurements \cite{Mak2010}  for $E_x > $ 6 MeV where a broad resonance structure has been observed for 6 MeV $< E_x <$ 10 MeV and interpreted as an E1 pygmy dipole resonance. This is consistent with our data (Fig. \ref{gsf139}) where the $\gamma$SF exhibits a broad feature for 6 MeV $< E_x < $ 9 MeV. 
Overall the three spin distributions from the HFB + Comb., BSFG1 + CT, and BSFG2 + CT models yield very similar $\gamma$SFs for each isotope (see Figs. \ref{gsf139} and \ref{gsf}). The  $\gamma$SF of $^{138}$La exhibits a low-energy enhancement for $E_{\gamma}$ $<$ 2 MeV, (Fig. \ref{gsf} (a)) for all tested spin-distributions. For $^{139}$La the strength function (Fig. \ref{gsf139}) could not be extracted for $E_{\gamma}$ $\leq$ 1.7 MeV due to non-statistical (discrete) features in the first-generation matrix. However,  it is obvious that the $\gamma$SFs of $^{139}$La exhibits a plateau behavior for $E_{\gamma} $ $<$ 3 MeV, similar to $^{138}$La which may be indicative of the development of a low-energy up-bend at energies below the measurement limit. A similar plateau structure is also observed in the $\gamma$SF of $^{140}$La for $E_{\gamma}$ $\leq$ 3 MeV (Fig. \ref{gsf}(b)) but no clear enhancement can be identified within the available $E_{x}$ range.  

The low-energy enhancement has been a puzzling feature since its first observation in $^{56, 57}$Fe \cite{voinov}. Its existence was independently confirmed using a different experimental and analytical technique in $^{95}$Mo \protect\cite{Wiedeking12} which triggered the study into the consistency of this feature with several $\gamma$SF models \cite{Kritcka2016}. Experimentally, the composition of the enhancement remains unknown, although it has been shown to be due to dipole transitions \cite{LarsenPRL2013, LarsenIoP2016}. Three theoretical interpretations have been brought forward to explain the underlying mechanism. According to Ref. \cite{Sch2013} this low-energy structure is due to M1 transitions resulting from a reorientation of spins of high-$j$ nucleon orbits, or due to 0$\hbar\omega$ M1 transitions \cite{Alex2014}. It has also been suggested that the up-bend could be of E1 nature due to single particle transitions from quasi-continuum to continuum levels \cite{Lit2013}. 

The emergence of the low-energy enhancement in the La isotopes is interesting and unexpected due to its prior non-observation for A $\geq$ 106 nuclei \cite{Larsen2013}. The appearance of this structure in La suggests that it is not confined to specific mass regions but may be found across the nuclear chart, an assumption that has recently received support through its observation in $^{151,153}$Sm \cite{Simon2016}.

\begin{figure}
	\centering
	\includegraphics[scale=0.46]{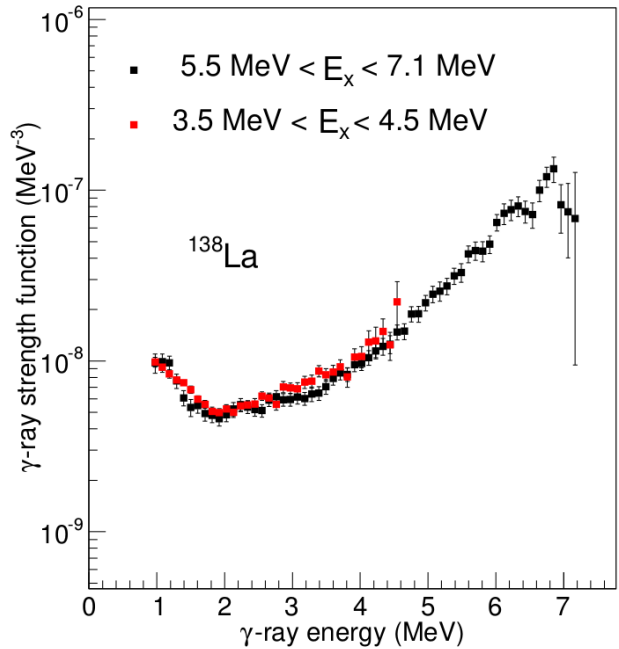}
	\caption{(color online) The $\gamma$-ray strength function of $^{138}$La extracted for two different excitation energy regions, and normalized with the HFB + Comb. spin distribution.}
	\label{ex_regions}
\end{figure}

The Brink hypothesis \cite{w} states that the $\gamma$SF of collective excitations is independent of the properties of initial and final nuclear states and only exhibits an $E_{\gamma}$ dependence. The validity of the Brink Hypothesis was experimentally verified for $\gamma$-ray transitions between states in the quasi-continuum \cite{guttormsen2016}. The independence of the set of quantum states from which the enhancement is extracted was confirmed for $^{138}$La where two non-overlapping $E_{x}$ regions have been independently used to measure the  $\gamma$SF, as shown in Fig. \ref{ex_regions}. It is apparent that the overall shape of the $\gamma$SF is indeed very similar for both excitation energy regions.    

\begin{figure}[h!]
	\centering
	\includegraphics[scale=0.58]{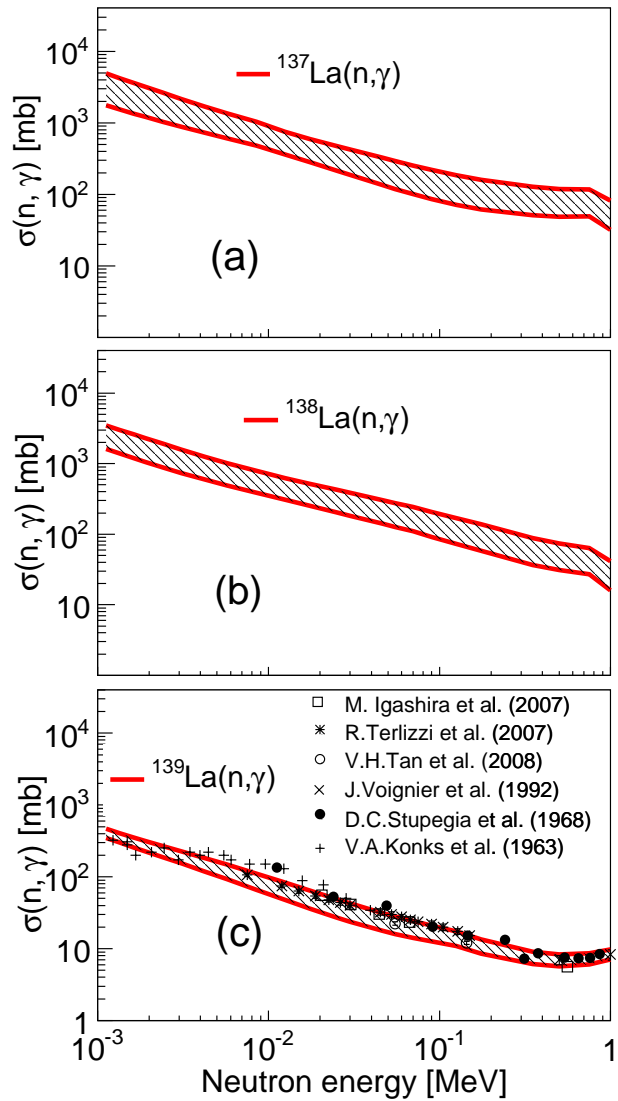}
	\caption{(color online) Calculated $^{137}$La$(n, \gamma)$, $^{138}$La$(n, \gamma)$ and $^{139}$La$(n, \gamma)$ cross-sections calculated with the TALYS reaction code using the measured NLDs and $\gamma$SFs as inputs. The $^{139}$La$(n, \gamma)$ cross-sections (c) are compared to available data from neutron-time of flight measurements (black data points) \cite{MIgashira2007,rTer2007,vhTan2008,JVoignier1992,DCStupegia1968,VAKonks1963}. The red lines indicate the upper and lower limits of the calculated cross sections.}
	\label{n_gamma_cross}
\end{figure}

\begin{table*}[htb]
	
	\caption{Astrophysical Maxwellian-averaged cross-sections.} 
	\begin{tabular}{lcccccccccccc}
		\hline
     	\hline
     	\\
		Reaction  &  & $(n, \gamma)$$^{138}$La & &  $(n, \gamma)$$^{139}$La &  & $(n, \gamma)$$^{140}$La &  & $(n, \gamma)$$^{138}$La  &   & $(n, \gamma)$$^{139}$La &   & $(n, \gamma)$$^{140}$La      \\
		
		Temperature (keV) &  &  30  &   &  30  &  &  30  & &   215  &  &   215 & &   215    \\
		
		MACS (mb) &  &  277.5$\pm$101  &  &  298$\pm$81 &  &  30.5 $\pm$6   & &   86 $\pm$34 &  & 26.5$\pm$10  & &  8.5 $\pm$2   \\
		
		\hline 
		\hline
	\end{tabular}
	\\
	\label{tab:nldpara}
	
	\label{table2}
\end{table*}

The presence of the low-energy enhancement in the A$\sim$ 140 region emphazises the need for systematic measurements to explore the extent and persistence of this feature, not only for nuclei near the line of $\beta$ stability but also for neutron-rich nuclei where the enhancement is expected to have significant impact on $r$-process reaction rates \cite{Larsen2010}. Establishing its electromagnetic character will also improve our understanding of the underlying physical mechanism of the enhancement and should be a priority for future measurements.   

The calculated NLDs using different models for the spin distribution (Figs. \ref{nld_La}, \ref{nld_LaC}, and \ref{nld_LaB}) are in good agreement with experimental data for all measured $E_x$ and for all La isotopes. The measured $\rho(E_{x})$ for $^{138,139,140}$La have very similar slopes, but are reduced for $^{139}$La compared to $^{138,140}$La. This behavior is due to odd-odd $^{138,140}$La nuclei having one extra degree of freedom that generates an increase in $\rho(E_{x})$ compared to odd-even $^{139}$La. The horizontal difference between NLDs of odd-odd and odd-even nuclei has been related to the pair gap parameter, while the vertical difference is a measure of entropy excess for the quasiparticle \cite{Moretto2015}. The constant temperature behavior of the NLDs (above the pair-breaking energy)  is a consistently observed feature \cite{GuttormsenNLD}, that is also confirmed by the HFB + Comb predictions, and has been interpreted as a first-order phase transition \cite{Moretto2015}.

According to the Hauser-Feshbach formalism \cite{Hauser1952} implemented in the TALYS code \cite{Kon2008}, the $^{A-1}$X$(n, \gamma)^{A}$X cross-sections are proportional to the $\gamma$-ray transmission coefficient, $\mathcal{T}(E_{\gamma})$, of a compound nucleus $^{A}$X.  This $\mathcal{T}(E_{\gamma})$ can in turn be determined from $\rho(E_{x}, J^{\pi})$ and $f(E_{\gamma})$, obtained from our measurement, and from which the $^{137}$La$(n, \gamma)$, $^{138}$La$(n, \gamma)$ and $^{139}$La$(n, \gamma)$ cross sections (see Fig. \ref{n_gamma_cross}) were computed. The statistical uncertainties of the experimental NLDs and $\gamma$SFs have been modified to include uncertainties in $D_{0}$ and $\langle   \varGamma_{\gamma}(S_{n},  J_{T}, \pi_{T})  \rangle$,  as discussed previously \cite{BVK2015}. These modifications to the uncertainties resulted in up to 69$\%$ and 34$\%$ uncertainties in the $\gamma$SFs and NLDs, respectively. For each La isotope we performed three cross-section calculations, in a consistent way, using the $\gamma$SFs and NLDs corresponding to the three adopted models (HFB + Comb., BSFG1 +CT  and BSFG2 + CT), resulting in very similar cross sections. The NLDs calculated with theoretical models were used in the excitation energy regions where they agree with the present experimental data, while our data points were interpolated and used in regions where they do not agree with calculated NLDs and discrete states (typically for $E_x<$ 2 MeV). In addition, the GSF was assumed to be of $E1$ character for these ($n$, $\gamma$) calculations. However, the effect of having the up-bend and pygmy resonance as M1 was also explored and this resulted in no change in the cross-sections. 

 Fig. \ref{n_gamma_cross}(c) shows the $^{139}$La$(n, \gamma)$ cross sections which are compared to the directly measured data  taken from \cite{MIgashira2007,rTer2007,vhTan2008,JVoignier1992,DCStupegia1968,VAKonks1963}. These are in excellent agreement and support the use of statistical nuclear properties to extract $(n,\gamma)$ cross sections, as previously discussed \cite{Laplace2016, Larsen2016,  Renstrom2016}. The comparison of the present cross-section data with those from direct measurements tests the reliability of using statistical decay properties to obtain ($n, \gamma$) cross sections and lends credibility to using this approach to also obtain reliable neutron-capture cross sections for $^{137}$La and $^{138}$La or for neutron-rich nuclei \cite{Spyrou2014, Liddick2016} for which no direct measurements are available

Futhermore, the normalized NLDs and GSFs were used to calculate the stellar Maxwellian-averaged cross-sections (MACS) at 30 and 215 keV which are the $s$- and $p$-process temperatures, respectively. These are shown in Tab. \ref{table2} for the $^{137}$La$(n, \gamma)$, $^{138}$La$(n, \gamma)$ and $^{139}$La$(n, \gamma)$ reactions. The present MACS for $^{137}$La$(n, \gamma)$ and $^{138}$La$(n, \gamma)$ are lower than those that were reported in \cite{BVK2015} by up to a factor of 2. This is due the newly determined $\gamma$SFs that are correspondingly lower than the previously at $E_{\gamma}$ $<$ 5 MeV due to the different normalization parameters. Nonetheless, for $^{138}$La at 215 keV the destructive $^{137}$La$(n, \gamma)$ MACS are three times the MACS of the producing reaction $^{138}$La$(n, \gamma)$. From these cross sections, it can be deduced~\cite{Goriely2001} that the synthesis of $^{138}$La through photodisintegration processes cannot be efficient enough to reproduce observed abundances, which is consistent with our previous results \cite{BVK2015}.

\section{Summary}
The NLDs and $\gamma$SF of $^{138,139,140}$La have been measured below $S_{n}$ using the Oslo Method. Three spin distributions, calculated with HFB + Comb. and the FG Model with two spin cut-off parameters,  were used for each La isotope for the normalization of these statistical nuclear properties. The NLDs were further compared with theoretical level densities obtained with HFB + Comb. and CT + FG approaches and are in reasonable agreement with the data. The excitation-energy independence of the low-energy enhancement  of $^{138}$La  has been verified in two different $E_{x}$ regions of the quasi-continuum which is consistent with the Brink hypothesis. Furthermore, the $\gamma$SFs of $^{139,140}$La are suggestive of the development of this  low-energy structure as well. None of the considered spin distributions, used for the normalization, can unambiguously eliminated it. The $^{137,138,139}$La$(n , \gamma)$ cross sections have been computed with the Hauser-Feshbach Model using consistently the NLDs and $\gamma$SFs data which are based on three distinct spin distributions. The $^{139}$La$(n , \gamma)$  cross sections were compared to available data and found to be in excellent agreement, giving confidence in the approach to obtain $(n , \gamma)$  cross sections from NLDs and $\gamma$SFs. The new MACSs calculated at 215 keV, for $^{138}$La$(n, \gamma)$ and $^{137}$La$(n, \gamma)$ reactions, confirm the underproduction of $^{138}$La in the $p$-proces.

\acknowledgments The authors would like to thank J. C. M\"uller, A. Semchenkov, and J. C. Wikne for providing excellent beam quality throughout the experiment and N.Y. Kheswa for manufacturing the target. This material is based upon work supported by the National Research Foundation of South Africa under grant nos. 92789 and 80365,  by the Research Council of Norway, project grant nos. 205528, 213442, and 210007, by US-NSF grants PHY-1204486 and PHY-1404343, by the US Department of Energy under contract no. DE-AC52-07NA27344, and the Department of Energy National Nuclear Security Administration under Award Number DE-NA0000979 through the Nuclear Science and Security Consortium. S.G. grants the support of the F.R.S.-FNRS. A.C.L. acknowledges funding from the Research Council of Norway, project grant no. 205528 and from ERC-STG-2014 Grant Agreement no. 637686. G.M.T gratefully acknowledges funding of this research from the Research Council of Norway, Project Grant no. 222287.

\end{document}